\documentclass[12pt]{article}
\makeatletter
\newcommand{\singlespacing}{\let\CS=\@currsize\renewcommand{\baselinestretch}{1}\tiny\CS}
\usepackage{epsfig}
\usepackage{lscape}
\usepackage{rotating}
\begin{document}
\baselineskip=24pt
\parskip = 10pt
\def \qed {\hfill \vrule height7pt width 5pt depth 0pt}
\newcommand{\ve}[1]{\mbox{\boldmath$#1$}}
\newcommand{\IR}{\mbox{$I\!\!R$}}
\newcommand{\1}{\Rightarrow}
\newcommand{\bs}{\baselineskip}
\newcommand{\esp}{\end{sloppypar}}
\newcommand{\be}{\begin{equation}}
\newcommand{\ee}{\end{equation}}
\newcommand{\beanno}{\begin{eqnarray*}}
\newcommand{\inp}[2]{\left( {#1} ,\,{#2} \right)}
\newcommand{\eeanno}{\end{eqnarray*}}
\newcommand{\bea}{\begin{eqnarray}}
\newcommand{\eea}{\end{eqnarray}}
\newcommand{\ba}{\begin{array}}
\newcommand{\ea}{\end{array}}
\newcommand{\nno}{\nonumber}
\newcommand{\dou}{\partial}
\newcommand{\bc}{\begin{center}}
\newcommand{\ec}{\end{center}}
\newcommand{\2}{\subseteq}
\newcommand{\cl}{\centerline}
\newcommand{\ds}{\displaystyle}
\def\refhg{\hangindent=20pt\hangafter=1}
\def\refmark{\par\vskip 2.50mm\noindent\refhg}

\title{\sc Simple Step-Stress Models with  a Cure Fraction}

\author{Nandini Kannan\footnote{Division of Mathematical Sciences, National Science Foundation, Arlington, VA 22230.  The research of Nandini Kannan was supported by the NSF IR/D program.  However, any opinion, finding, and conclusions and recommendations expressed in this material are those of the author and do not necessarily reflect the views of the National Science Foundation} \& Debasis Kundu\footnote{Department of Mathematics and
Statistics, Indian Institute of Technology Kanpur, Kanpur, Pin 208016, India}}

\date{}
\maketitle

\begin{abstract}
In this article, we consider models for time-to-event data  obtained 
from experiments in which stress levels are altered at intermediate
stages during the observation period.  
These experiments, known as step-stress tests, belong to the larger class
of accelerated tests used extensively
in the reliability literature.
 The analysis of data from step-stress tests largely relies on the popular cumulative exposure model.    
However, despite its simple form, the utility of the model is limited, as it is assumed that the hazard function of the 
underlying distribution is discontinuous at the points at which the stress levels are changed, which may not be very reasonable.  Due to this deficiency, Kannan et al. \cite{KKNT:2010}
introduced the cumulative risk model, where the hazard function is continuous.  In this paper we propose a class of parametric 
models based on the cumulative risk model assuming the underlying population contains long-term survivors or `cured' fraction.  An EM algorithm to
compute the maximum likelihood estimators of the unknown parameters is proposed.   This research is motivated by a study on altitude decompression sickness.  The performance of different parametric models will be evaluated using data from this study.

\end{abstract}


\newpage

\section{\sc Introduction}
In many applications in the physical, environmental, and biomedical sciences, the experimental conditions
may change during the exposure period.  This research is motivated by a study on altitude decompression sickness (DCS), a condition  caused by a decrease in pressure and lower oxygen levels at high altitudes.  Mountaineers and travelers to high mountainous regions frequently experience symptoms that include nausea, headaches, dizziness and fatigue.  Decompression sickness is also observed in deep-sea divers and mine workers.  The change in atmospheric pressure causes nitrogen to come out of solution and form bubbles in tissues and the bloodstream.  Symptoms of DCS range from mild joint pain and swelling to dizziness, and in extreme cases death.  
To avoid/ decrease
symptoms of decompression sickness, mountaineers are advised to acclimatize to the altitude conditions and perform gradual staged ascents.  Deep-sea divers are provided with tables that specify the length of time at each stage in the ascent from depth (decompression stops)  to avoid the "bends".  This
staged ascent can delay or eliminate the onset of symptoms by allowing nitrogen levels  to gradually dissipate.  In the staged ascent, the experimental conditions (in this case, altitude) are changed throughout the exposure period.  

Greven et al. \cite{Greven:2004} described an experiment designed to examine the effects of water contamination on fish, in
particular on their swimming performance.  The underlying hypothesis was that fish exposed to toxins
may exhibit a lower threshold for fatigue.  Fatigue was induced by increasing the stress, in this case the water velocity, at fixed time points during the experiment.

The two examples described above are known as "step-stress tests" in the reliability literature.  Step-stress tests belong to the larger class of accelerated tests that have been extensively used in life-testing
experiments.  As products become more reliable, standard life-testing experiments often fail to provide adequate data on failure times, resulting in poor estimates of reliability.
Accelerated testing allows the researcher to overcome this problem by subjecting  the experimental units or
items to higher than normal levels of stress.  The increase in stress reduces the time to failure by accelerating the degradation or deterioration of the item.  Using a model that relates the stress level
to the failure distribution, properties of the underlying failure distribution
under normal operating conditions may be obtained.
In a standard step-stress experiment, all individuals or items are subject to an initial stress level.  The
stress is gradually increased at pre-specified times during the experiment.  The stress factor may refer to the
dose of a drug, incline/speed of a treadmill, temperature, voltage, or pressure.

The most commonly used model for step-stress experiments is the cumulative exposure model (CEM) introduced by Sedyakin  \cite{Sedy:1966}.  The CEM and its variants have been studied by a number of authors including Bai et al. \cite{BKL:1989}, Balakrishnan \cite{Bala:2009}, Balakrishnan et al. \cite{BKNK:2007},
Gouno and Balakrishnan \cite{GB:2001}, Nelson \cite{Nelson:1980, Nelson:1990}, Xiong \cite{Xiong:1998},
Xiong and Milliken \cite{XM:1999}, Khamis and Higgins \cite{KH:1998}, Ismail \cite{Ismail:2016},  Kateri and Kamps 
\cite{KK:2015} and Sha and Pan \cite{SP:2014}.
The formulation of the CEM provides a relationship between the lifetime distribution of an experimental unit at one stress level and the life-time
distribution of the unit at a second stress level.  As a consequence of this assumption, the underlying hazard function is discontinuous at the time point at which the stress level changes. This implies that the effect of
a change in stress level is instantaneous, which may not be realistic in practice.

Kannan et al. \cite{KKB:2010} proposed a new simple step-stress model assuming that there is a latency  or lag
period $\delta$  before the effect of the change in stress level is observed.  The cumulative risk model (CRM), proposed by the authors, assumes that the hazard function for the step stress experiment is continuous.  The CEM and the time transformed Khamis and Higgins
\cite{KH:1998} model may be obtained as limiting cases of the CRM.
 Kannan et al. \cite{KKB:2010} developed maximum likelihood and
least squares estimators for the parameters of the CRM assuming  exponential lifetime distributions at the two stress levels.  Beltrami \cite{Beltrami:2011} extended the CRM to account for competing causes of failure assuming exponential and Weibull lifetime distributions.

In recent years, the development of new drugs and treatment regimens has resulted in patients living longer
with certain diseases.  In some cases, patients show no recurrence of a disease, i.e are viewed as being permanently cured.  These patients are referred to as "cured'' or long term survivors, while the remaining patients are referred to as "susceptible".  Traditional parametric survival models such as Weibull or Gamma do not account 
for the probability of cure.  Although subtle, it is important to distinguish between the concepts of censoring and cure.  Note 
that censoring refers to a subject who does not fail within the monitoring time window of a particular item, where as cure 
refers to one who will not fail within any reasonable time window.  Clearly, the later is an abstraction, as we never observe a 
cure due to a finite monitoring time.  Still estimating the probability of such an outcome is a important issue, see for example 
Tsodikov \cite{Tsodikov:1998}.  In the context of reliability or life-testing applications, cure refers to items that are resistant and
not affected by the stress factor.  It can happen even in a step-stress experiment also.  For example, in the DCS study it has been
observed, see for example Kannan et al. \cite{KRP:1998},  that even in the increased stress level some of the experimental units are immune to the stress factors and they do not
fail at the higher stress level.  They can be treated as the long term survivors.

The goal of this article is to develop a cumulative risk model assuming the underlying population includes both `cured' and susceptible individuals.    We consider different
parametric distributions to model the lifetimes of susceptible individuals.    
In Section 2, we define the generalized cumulative risk model and discuss three special cases.  In Section 3, using the standard mixture model formulation, the generalized CRM is adapted to include a cured fraction.  The probability of being cured is modeled as a function of covariates using a logistic link.  Section 4 describes an 
EM algorithm for obtaining the MLEs of the parameters and associated likelihood ratio tests.  In Section 5, we fit different parametric 
models to the data on altitude decompression sickness and evaluate the goodness of fit of these models.  Conclusions and future research directions are discussed in Section 6.

\section{\sc Generalized Cumulative Risk Model}

Consider a life-testing experiment wherein the stress is changed only once during the exposure duration.  We assume that all $n$
individuals or items under study are exposed to an initial stress level $x_1$.  The items are continuously
monitored, and at some pre-specified time $\tau_1$, the stress level is changed (increased) to $x_2$.  We assume that there is a latency
period $\delta$ before the effects of the increased stress are observed.   

We assume that the hazard function is continuous for all $t > 0$ and is increasing in the interval $[\tau_1, \tau_2 ]$, where $\tau_2 = \tau_1 + \delta$.  
  The piecewise hazard function for the Generalized Cumulative Risk Model has the following form;
\be
h_0(t) = \left \{ \matrix{h_{01}(t) & \hbox{if} & 0 < t \le \tau_1  \cr
   &  &   \cr
a + b t & \hbox{if} & \tau_1 < t < \tau_2   \cr
& &  \cr
h_{03}(t) & \hbox{if} & \tau_2 \le t < \infty}.   \right .   \label{2.1}
\ee
Here $h_{01}(t) > 0$ and $h_{03}(t) > 0$ are continuous functions for $t > 0$, and 
\be
\int_0^{\infty} h_0(t) dt = \infty.    \label{cond-hazard}
\ee
The parameters $a$ and $b$ are such
that the hazard function $h_0(t)$ is a continuous function for $t > 0$.  Therefore, $a$ and $b$ satisfy the following conditions:
\bea
a + b \tau_1 & = & h_{01}(\tau_1),   \label{2.2}   \\
a + b \tau_2 & = & h_{03}(\tau_2).    \label{2.3}
\eea
Note that, although for mathematical tractability, we assume the hazard function to be linear in the interval $[\tau_1,\tau_2]$, other smooth functions may be considered.  

The generalized  cumulative risk model, as defined through the hazard function (\ref{2.1}), is
an extension of the cumulative risk model proposed by Kannan et al. \cite{KKB:2010}.  The authors  considered the case when $h_{01} = \lambda_1$ and $h_{03}(t) = \lambda_2$, i.e.
exponential distributions with different scale parameters.  

When 
$\tau_2 \downarrow \tau_1 = \tau$ (say), then (\ref{2.1}) takes the form
\be
h_0(t) = \left \{ \matrix{h_{01}(t) & \hbox{if} & 0 < t \le \tau  \cr
&  & \cr
h_{03}(t) & \hbox{if} & \tau < t < \infty.  \cr}  \right .   \label{cem-hazard}
\ee
Equation (\ref{cem-hazard}) is the hazard function for  the cumulative exposure model in the case of exponential
hazards (see Balakrishnan \cite{Bala:2009} or Kateri and Balakrishnan \cite{KB:2008}), and the hazard function  of the model proposed by 
Khamis and Higgins \cite{KH:1998} in the case of Weibull hazards, when the two shape parameters are equal.  It can also be observed as 
an generalization of the tampered failure rate (TFR) model of Bhattacharyya and Soejoeti (1989).  In the TFR model $h_{01}(t)/h_{03}(t)$ 
is a constant.

From (\ref{2.1}), we obtain the
cumulative hazard function as
\be
H_0(t) = \left \{ \matrix{H_{01}(t) & \hbox{if} & 0 < t \le  \tau_1   \cr
&  &   \cr
H_{02}(t) & \hbox{if} & \tau_1 < t < \tau_2   \cr
&  &  \cr
H_{03}(t) & \hbox{if} & \tau_2 \le t < \infty  \cr},  \right .   \label{2.4}
\ee
where
\beanno
H_{01}(t) & = & \int_0^t h_{01}(u) du   \\
H_{02}(t) & = & H_{01}(\tau_1) + a(t-\tau_1) +\frac{b}{2} (t^2 - \tau_1^2)  \\
H_{03}(t) & = & H_{02}(\tau_2) + \int_{\tau_2}^t h_{03}(u) du.
\eeanno
Therefore, the survival function is given by
\be
S_0(t) = \left \{ \matrix{e^{-H_{01}(t)} & \hbox{if} & 0 < t \le  \tau_1   \cr
&  &   \cr
e^{-H_{02}(t)} & \hbox{if} & \tau_1 < t < \tau_2   \cr
&  &  \cr
e^{-H_{03}(t)} & \hbox{if} & \tau_2 \le t < \infty.  \cr}   \right .   \label{2.5}
\ee
Note that due to the condition (\ref{cond-hazard}), (\ref{2.5}) is a proper survival function.  The associated probability 
density function (PDF) is
\be
f_0(t) = \left \{ \matrix{f_{01}(t) & \hbox{if} & 0 < t \le \tau_1   \cr
&  &   \cr
f_{02}(t) & \hbox{if} & \tau_1 < t < \tau_2   \cr
&  &  \cr
f_{03}(t) & \hbox{if} & \tau_2 \le t < \infty,  \cr}  \right .
\ee
where
\be
f_{01}(t)  =   h_{01}(t) \ e^{-H_{01}(t)}, \ \ f_{02}(t) = (a+bt) \ e^{-H_{02}(t)}, \ \
f_{03}(t) = h_{03}(t) \ e^{-H_{03}(t)}.
\ee

\subsection{Special Cases}

In this subsection, we describe three special cases of the generalized cumulative risk model.

\noindent {\sc Weibull Model:} The Weibull distribution is one of the most widely used parametric distribution in the survival analysis and reliability literature to analyze lifetime data.  The PDF of a two-parameter Weibull
distribution has the following form:
\be
f(t) = \alpha \lambda t^{\alpha-1} e^{-\lambda t^{\alpha}}; \ \ \  x > 0.
\ee
Here $\alpha > 0$ and $\lambda > 0$ are the shape and scale parameters, respectively.  

Therefore, we have
\be
h_{01}(t)  =  \alpha_1 \lambda_1 t^{\alpha_1-1},  \ \ \ \
h_{03}(t)  =  \alpha_2 \lambda_2 t^{\alpha_2-1}
\ee
and
\beanno
H_{01}(t)  & =  & \lambda_1 t^{\alpha_1}, \\
H_{02}(t)  & =  & \lambda_1 \tau_1^{\alpha_1} + a(t-\tau_1) +\frac{b}{2} (t^2 - \tau_1^2)  \\
H_{03}(t)  & =  & \lambda_1 \tau_1^{\alpha_1} + a(\tau_2-\tau_1) +\frac{b}{2} (\tau_2^2 - \tau_1^2) +
(\lambda_2 \tau_2^{\alpha_2} - \lambda_2 t^{\alpha_2}).
\eeanno

\noindent{\sc Linear Failure Rate Distribution:} The probability density function of a two-parameter linear failure
rate distribution for $\alpha > 0$ and $\lambda > 0$, is given by
\be
f(t) = (\alpha + \lambda t) e^{-\alpha t - \frac{\lambda t^2}{2}}.
\ee
In this case,
\be
h_{01}(t) = \alpha_1 + \lambda_1 t, \ \ \ \ h_{03}(t) = \alpha_2 + \lambda_2 t
\ee
and
\beanno
H_{01}(t)  & =  & \alpha_1 t + \frac{\lambda_1 t^2}{2}, \\
H_{02}(t)  & =  & \alpha_1 \tau_1 + \frac{\lambda_1 \tau_1^2}{2} + a(t-\tau_1) +\frac{b}{2} (t^2 - \tau_1^2)  \\
H_{03}(t)  & =  & \alpha_1 \tau_1 + \frac{\lambda_1 \tau_1^2}{2} + a(\tau_2-\tau_1) +\frac{b}{2}
(\tau_2^2 - \tau_1^2) +
\alpha_2(t - \tau_2) + \frac{\lambda_2}{2} (t^2 - \tau_2^2).
\eeanno

\noindent {\sc Generalized Exponential Distribution:} This lifetime model was introduced by Gupta and Kundu
\cite{GK:1999} as an alternative to the gamma and Weibull distributions.  The two-parameter generalized
exponential distribution has the following PDF:
\be
f(t) = \alpha \lambda e^{-\lambda t} (1 - e^{-\lambda t})^{\alpha -1}; \ \ \ \ t > 0,
\ee
where $\alpha > 0$ and $\lambda > 0$  are the shape and scale parameters, respectively.  

In this case
\be
h_{01}(t) = \frac{\alpha_1 \lambda_1 e^{-\lambda_1 t} (1 - e^{-\lambda_1 t})^{\alpha_1-1}}
{1 - (1 - e^{-\lambda_1 t})^{\alpha_1}}
, \ \ \ \ h_{03}(t) = \frac{\alpha_2 \lambda_2 e^{-\lambda_2 t} (1 - e^{-\lambda_2 t})^{\alpha_2-1}}
{1 - (1 - e^{-\lambda_2 t})^{\alpha_2}}
\ee
and
\beanno
H_{01}(t)  & =  & -\ln \left ( 1 - (1 - e^{-\lambda_1 t})^{\alpha_1} \right ),  \\
H_{02}(t)  & =  & -\ln \left ( 1 - (1 - e^{-\lambda_1 \tau_1})^{\alpha_1} \right ) + a(t-\tau_1) +\frac{b}{2} (t^2 - \tau_1^2)  \\
H_{03}(t)  & =  & -\ln \left ( 1 - (1 - e^{-\lambda_1 \tau_1})^{\alpha_1} \right ) + a(\tau_2-\tau_1) +
\frac{b}{2} (\tau_2^2 - \tau_1^2) \\
&  & \ \ \
-\ln \left ( 1 - (1 - e^{-\lambda_2 t})^{\alpha_2} \right ) -
\ln \left ( 1 - (1 - e^{-\lambda_2 \tau_2})^{\alpha_2} \right ).
\eeanno
For all three cases, let ${\ve \Theta} = (\alpha_1, \alpha_2, \lambda_1, \lambda_2)$ denote the parameter vector.

\section{CRM with Long-Term Survivors}

Extensive work has been done in formulating parametric and non-parametric survival models incorporating a cured fraction, a 
non-zero tail probability of the survival function, starting with the seminal paper by Boag \cite{Boag:1949}, who proposed a two 
component mixture model to analyze breast cancer data.
The population of interest may be regarded as a mixture of cured and susceptible 
individuals/ items.  In this model, the survival function for the entire population can be written as
$$
P(T > t) = S_p(t) = \pi + (1-\pi)S(t),
$$
where $\pi = S_p(+\infty)$ is the `cured fraction', and $S(t)$ with $S(+\infty)$ = 0 is the proper survival function for the 
`non-cured' or the susceptible population. Since, then there has been extensive research on the cure rate model, primarily in the survival
analysis literature.  Interested readers may refer to the classical book by Maller and Zhou \cite{MZ:1996} or
Yu et al. \cite{YTCF:2004}, Cancho and Bolfarine \cite{CB:2001},
Chen et al. \cite{CIS:1999}, Gamel et al. \cite{GMR:1999},
Kannan et al. \cite{KKNT:2010}, and the references cited therein.

The cumulative risk model can be extended to incorporate the cured fraction that represents individuals/items resistant to the stress factors.
We assume that the susceptible population has a survival function $S_0(\cdot)$
as defined in (\ref{2.5}).  
The probability that an individual is cured is assumed to depend on a set of $s$ covariates $z_{1}, \ldots, z_{s}$.  Let ${\ve z} = (1, z_{1}, \ldots, z_{s})$.  For each individual under study with covariate vector ${\ve z}$,
we define an indicator random variable $\Delta({\ve z})$, with $\Delta({\ve z})$ = 1,
when the subject is susceptible, and $\Delta({\ve z})$ = 0, when the subject is cured.  It is 
assumed that
$$
P(\Delta({\ve z}) = 0) = p({\ve \beta}, {\ve z}) =
\frac{e^{{\ve \beta}'{\ve z}}}{1 + e^{{\ve \beta}'{\ve z}}} \ \ \ \ \
\hbox{and} \ \ \ \ \ \
P(\Delta({\ve z})=1) = 1-p({\ve \beta}, {\ve z}) =
\frac{1}{1 + e^{{\ve \beta}'{\ve z}}}.
$$
Here $\ve{\beta}$ is an $s+1$ dimensional vector of coefficients and  $\beta_{0}$ represents the intercept term.
Let $S(t;{\ve \Theta}, {\ve \beta}, {\ve z})$ denote the survival function of an individual with covariate vector ${\ve z}$.  We have 
\be
S(t;{\ve \Theta}, {\ve \beta}, {\ve z}) = p({\ve \beta}, {\ve z}) + (1-p({\ve \beta}, {\ve z})) S_0(t; {\ve \Theta}),   \label{2.6}
\ee
where $S_0(t; {\ve \Theta})$ has the form of the cumulative risk model defined in (\ref{2.5}). 

It may be noted that we have linked the covariates only with the cure fraction and not with the survival function 
$S_0(t; {\ve \Theta})$.  It may be argued that would provide more general model.  Although, it may be true but it raises some
questions regarding choosing the proper covariates for two different purposes, and their identifiability issues, see also 
Kannan et al. \cite{KKNT:2010}.  Due to which it has been avoided.

\section{\sc Statistical Inference}

\subsection{\sc Maximum Likelihood Estimation}

In this section, we propose an EM algorithm to compute  the maximum likelihood estimators (MLEs) of the unknown parameters, namely,
${\ve \Theta}$ and ${\ve \beta}$.   
At the initial stage, we assume that $\delta$, or equivalently,  $\tau_2 = \tau_1 + \delta$ is known.
We will then describe a procedure to estimate $\delta$.  

The observed failure times from the step-stress experiment are denoted by 
\be
0 < t_1 < \cdots < t_{n_1}
< \tau_1 < t_{n_1+1} < \cdots < t_{n_1+n_2} < \tau_2 <
t_{n_1+n_2+1} < \cdots \le t_{n_1+n_2+n_3}.   \label{3.1}
\ee
Here $m = n_1+n_2+n_3$ represents the total number of failures observed.  The remaining observations 
\be
t_{n_1+n_2+n_3+1} < \ldots < t_{n_1+n_2+n_3+n_4},   \label{3.2}
\ee
where $n_4 = n - m$, are censored.   With each failure/censor time $t_i$, we have
an associated covariate vector ${\ve z}_i$.

Let $I_1 = \{1, \cdots, n_1\}$, $I_2 = \{n_1+1, \cdots, n_1+n_2\}$, $I_3 = \{n_1+n_2+1, \cdots, n_1+n_2+n_3\}$
and $I_4 = \{n_1+n_2+n_3+1, \cdots, n_1+n_2+n_3+n_4\}$.  Here $I_4$ denotes the set of censored observations.
Based on the observations $\{(t_i, {\ve z}_i); i = 1, \ldots, n\}$, the log-likelihood function can be written as
\be
l({\ve \Theta}, {\ve \beta}) = l_1({\ve \Theta}, {\ve \beta}) + l_2({\ve \Theta}, {\ve \beta}),
\label{ll}
\ee
where
$$
l_1({\ve \Theta}, {\ve \beta})  =  \sum_{i \in I_1 \cup I_2 \cup I_3} \ln(1-p({\ve \beta}, {\ve z}_i)) + \sum_{i \in I_1} \ln
f_{01}(t_i; {\ve \Theta}) + \sum_{i \in I_2} \ln f_{02}(t_i; {\ve \Theta}) +
\sum_{i \in I_3} \ln f_{03}(t_i; {\ve \Theta}),
$$
and
$$
l_2({\ve \Theta}, {\ve \beta})  =  \sum_{i \in I_4}  \ln \left \{p({\ve \beta}, {\ve z}_i) +
(1-p({\ve \beta}, {\ve z}_i)) S_0(t_i; {\ve \Theta}))  \right \}.
$$
The maximum likelihood estimators of ${\ve \Theta}$ and ${\ve \beta}$ can be obtained by maximizing the log-likelihood
function (\ref{ll}), and involves a $(4+(s+1))$-dimensional optimization
problem, where $s$ denotes the number of covariates.

The MLEs may also be obtained by treating the problem as a  missing data problem and using the EM algorithm.
For an individual with covariate vector ${\ve z}$ in $I_1 \cup I_2 \cup I_3$,
the associated random variable $\Delta({\ve z})$ takes the value 1.  For an individual in $I_4$,
$\Delta({\ve z})$ is unobserved.  Therefore, for these $n_4$ observations, the associated $\Delta({\ve z})$ values are treated as missing.

At the $k-th$ stage of the EM algorithm, let ${\ve \Theta}^{(k)}$ and
${\ve \beta}^{(k)}$ denote the estimates of the parameters 
${\ve \Theta}$ and ${\ve \beta}$, respectively.  We then compute the pseudo log-likelihood function  based on the missing
observations ('E'-step).  For a censored time $t$, two partially complete 'pseudo
observation' of the form $(t, w_1(t|\Theta^{(k)}, {\ve \beta}^{(k)},{\ve z}))$ and
$(t, w_2(t|{\ve \Theta}^{(k)}, {\ve \beta}^{(k)}, {\ve z}))$ are constructed.  Here,
$w_1(t|{\ve \Theta}^{(k)}, {\ve \beta}^{(k)}, {\ve z})$ and $w_2(t|{\ve \Theta}^{(k)},
{\ve \beta}^{(k)}, {\ve z})$
denote the conditional probabilities that the
individual with covariate ${\ve z}$ belongs to the cure and susceptible group, respectively, given that the individual has
survived till time $t$.  We have
$$
w_1(t|{\ve \Theta}^{(k)}, {\ve \beta}^{(k)},{\ve z}) = P(\Delta({\ve z}) = 0| T > t, {\ve \Theta}^{(k)})
$$
and
$$
w_2(t|{\ve \Theta}^{(k)}, {\ve \beta}^{(k)},{\ve z}) = P(\Delta({\ve z}) = 1| T > t, {\ve \Theta}^{(k)}).
$$
We have
\beanno
w_1(t|{\ve \Theta}^{(k)},{\ve \beta}^{(k)},{\ve z})  & = &
P(\Delta({\ve z}) = 0| T > t, {\ve \Theta}^{(k)},{\ve \beta}^{(k)}) \\
&  = &
\frac{P(T > t| \Delta({\ve z}) = 0, {\ve \Theta}^{(k)}, {\ve \beta}^{(k)})
\times P(\Delta({\ve z}) = 0|\beta^{(k)})}{P(T > t| {\ve \Theta}^{(k)}, {\ve \beta}^{(k)})} \\
& = &
\frac{p({\ve \beta}^{(k)}, {\ve z})}{p({\ve \beta}^{(k)},{\ve z}) +
(1-p({\ve \beta}^{(k)}, {\ve z})) S_0(t|{\ve \Theta}^{(k)})},
\eeanno
and
\beanno
w_2(t|{\ve \Theta}^{(k)},{\ve \beta}^{(k)},{\ve z})  & = &
P(\Delta({\ve z}) = 1| T > t, {\ve \Theta}^{(k)},{\ve \beta}^{(k)}) \\
&  = &
\frac{P(T > t| \Delta({\ve z}) = 1, {\ve \Theta}^{(k)}, {\ve \beta}^{(k)})
\times P(\Delta({\ve z}) = 1|\beta^{(k)})}{P(T > t| {\ve \Theta}^{(k)},{\ve \beta}^{(k)})} \\
& = &
\frac{(1-p({\ve \beta}^{(k)}, {\ve z})) S_0(t|{\ve \Theta}^{(k)})}{p({\ve \beta}^{(k)}, {\ve z}) +
(1-p({\ve \beta}^{(k)}, {\ve z})) S_0(t|{\ve \Theta}^{(k)})}.
\eeanno

Therefore, the 'pseudo log-likelihood' function, $L_{pseudo}({\ve \Theta}, {\ve \beta})$ based on the
missing observation is given by
\bea
l_{pseudo}({\ve \Theta}, {\ve \beta}|{\ve \Theta}^{(k)}, {\ve \beta}^{(k)}) & = &
l_1({\ve \Theta}, {\ve \beta}) \nonumber  \\
& & + \sum_{i \in I_4}
\left \{w_1(t_i|{\ve \Theta}^{(k)}, {\ve \beta}^{(k)}) \ln p({\ve \beta}, {\ve z}_i)  \right . \nonumber \\
& & \ \ \ \ \ \ \ \  +
\left . w_2(t_i|{\ve \Theta}^{(k)}, {\ve \beta}^{(k)}) \ln \left [ (1-p({\ve \beta}, {\ve z}_i))S_0(t_i|{\ve \Theta}) \right ] \right \}  \nonumber  \\
& = & \sum_{i \in I_1 \cup I_2 \cup I_3} \ln (1-p({\ve \beta}, {\ve z}_i))
+ \sum_{i \in I_4} w_1(t_i|{\ve \Theta}^{(k)}, {\ve \beta}^{(k)}) \ln p({\ve \beta}, {\ve z}_i) \nonumber
\\
&  & + \sum_{i \in I_4} w_2(t_i|{\ve \Theta}^{(k)}, {\ve \beta}^{(k)}) \ln (1 - p({\ve \beta}, {\ve z}_i) ) \nonumber  \\
&  & +
\sum_{i \in I_1} \ln f_{01}(t_i|{\ve \Theta}) + \sum_{i \in I_2} \ln f_{02}(t_i|{\ve \Theta}) \nonumber  \\
&  & +
\sum_{i \in I_3} \ln f_{03}(t_i|{\ve \Theta}) + \sum_{i \in I_4} w_2(t_i|{\ve \Theta}^{(k)}, {\ve \beta}^{(k)})
\ln S_0(t_i|{\ve \Theta})  \nonumber  \\
& =& g_1({\ve \beta}|{\ve \Theta}^{(k)}, {\ve \beta}^{(k)}) + g_2({\ve \Theta}|\Theta^{(k)}) \ \ \ \ (\hbox{say}),
\eea
where
\bea
g_1({\ve \beta}|{\ve \Theta}^{(k)}, {\ve \beta}^{(k)}) & = &
\sum_{i \in I_1 \cup I_2 \cup I_3} \ln (1-p({\ve \beta}, {\ve z}_i))
+ \sum_{i \in I_4} w_1(t_i|{\ve \Theta}^{(k)}, {\ve \beta}^{(k)}) \ln p({\ve \beta}, {\ve z}_i)
\nonumber   \\
& & \ \ \ \ \
+ \sum_{i \in I_4} w_2(t_i|{\ve \Theta}^{(k)}, {\ve \beta}^{(k)}) \ln (1 - p({\ve \beta}, {\ve z}_i) ),
\label{opt-1}
\eea
and
\be
g_2({\ve \Theta}|{\ve \Theta}^{(k)})  = g_3({\ve \Theta}) + \sum_{i \in I_4}
w_2(t_i|{\ve \Theta}^{(k)}, {\ve \beta}^{(k)}) \ln S_0(t_i|{\ve \Theta}).   \label{opt-2}
\ee
Here
\be
g_3({\ve \Theta})  =
\sum_{i \in I_1} \ln f_{01}(t_i|{\ve \Theta}) + \sum_{i \in I_2} \ln f_{02}(t_i|{\ve \Theta}) +
\sum_{i \in I_3} \ln f_{03}(t_i|{\ve \Theta}).   \label{g3}
\ee
The 'M'-step of the EM algorithm involves maximization of $\ds l_{pseudo}({\ve \Theta}|{\ve \Theta}^{(k)})$
 with respect to the unknown parameters.

Assuming a model with no covariates, and denoting $p({\ve \beta}, {\ve z})  = p$, the 
function (\ref{opt-1}) reduces to 
\be
g_1(p|{\ve \Theta}^{(k)}, p^{(k)})  =
m \ln (1-p)
+ \ln p \sum_{i \in I_4} w_1(t_i|{\ve \Theta}^{(k)}, p^{(k)})
+ \ln (1-p) \sum_{i \in I_4} w_2(t_i|{\ve \Theta}^{(k)}, p^{(k)}),
\ee
where
\be
w_1(t_i|{\ve \Theta}^{(k)}, p^{(k)})  =  \frac{p^{(k)}}{p^{(k)} + (1-p^{(k)})S_0(t_i|{\ve \Theta}^{(k)})}
\ee
and
\be
w_2(t_i|{\ve \Theta}^{(k)}, p^{(k)}) = \frac{(1-p^{(k)})S_0(t_i|{\ve \Theta}^{(k)})}{p^{(k)} + (1-p^{(k)})S_0(t_i|{\ve \Theta}^{(k)})}.
\ee
Therefore
\be
p^{(k+1)} = \frac{\sum_{i \in I_4} w_1(t_i|{\ve \Theta}^{(k)}, p^{(k)})}{n}.
\ee

If the model includes covariates, a Newton-Raphson type method or other optimization techniques may be used to maximize
the functions $g_1({\ve \beta}|{\ve \Theta}^{(k)}, {\ve \beta}^{(k)})$ and  
$g_2({\ve \Theta}|{\ve \Theta}^{(k)})$.

So far we have assumed that $\delta$ is known.  Since in practice $\delta$ is unknown we estimate $\delta$ by maximizing the
profile likelihood function with respect to $\delta$. Since it cannot be done analytically we have suggested to use grid search
method for this purpose.  For each $\delta$ we compute the MLEs of the unknown parameters as suggested above and obtain the 
corresponding likelihood value, and then choose that $\delta$ for which the likelihood value is maximum.  The detailed 
implementation procedure will be explained in the Data Analysis section.

\subsection {\sc Testing of Hypothesis}

In this section, we describe three hypotheses of interest and derive the corresponding likelihood ratio tests.  In all the 
cases the exact tests are not available, hence, we rely on the asymptotic tests.  Since the sample size is not very small, 
it is well known that the asymptotic tests work reasonably well.

\noindent {\sc Problem 1:} 
\be
{{{\cal{H}}}}_{01}: \alpha_1 = \alpha_2 \ \ \ \  \hbox{vs.} \ \ \ \ \ {{{\cal{H}}}}_{11}: \alpha_1 \ne \alpha_2.
\label{prob-1}
\ee
The hypothesis tests  the equality of the shape parameters of the corresponding
lifetime distributions at the two different stress levels.  Under the null hypothesis, the maximization
of $g_2({\ve \Theta}|{\ve \Theta}^{(k)})$ involves a three-dimensional optimization.
 Letting $l_{01}$ and $l_{11}$ denote
the maximized log-likelihood values under the null and alternative hypothesis, respectively, we have
\be
- 2(l_{01} - l_{11}) \sim \chi^2_1.
\ee

\noindent {\sc Problem 2:} 
\be
{{{\cal{H}}}}_{02}: \alpha_1 = \alpha_2 = 1 \ \ \ \  \hbox{vs.} \ \ \ \ \ {{{\cal{H}}}}_{12}: \alpha_1 = \alpha_2 \ne 1.
\label{prob-2}
\ee
The null hypothesis states that the lifetime distributions at the two different stress levels are exponential. Under the null hypothesis,  the maximization
of $g_2({\ve \Theta}|{\ve \Theta}^{(k)})$ involves a two-dimensional optimization.  Letting $l_{02}$ and $l_{12}$ (= $l_{11}$)
denote
the maximized log-likelihood values under the null and alternative hypothesis, respectively, we have 
\be
- 2(l_{02} - l_{12}) \sim \chi^2_1.
\ee

\noindent {\sc Problem 3:} 
\be
{{{\cal{H}}}}_{03}: \beta_1 = \beta_1 = \ldots = \beta_s = 0 \ \ \ \ \ \hbox{vs.} \ \ \ \
{{{\cal{H}}}}_{13}: \hbox{At least one of them is not 0}
\ee
This is a  test of the significance of the covariates.  In
the EM algorithm, the 'M'-step that involves the maximization of
$g_1(p|{\ve \Theta}^{(k)},p^{(k)})$ can be performed explicitly.  From the likelihood ratio test, it
follows that
\be
- 2(l_{03} - l_{13}) \sim \chi^2_s,    \label{test-3}
\ee
where $l_{03}$ and $l_{13}$ denote the maximized log-likelihood values under the null and alternative
hypothesis, respectively.

\noindent {\sc Problem 4:} 
\be
{{{\cal{H}}}}_{04}: p = 0 \ \ \ \ \ \hbox{vs.} \ \ \ \
{{{\cal{H}}}}_{14}: p > 0.
\ee
This test assesses whether or not the underlying population has a cure fraction.    Since
$0$ is a boundary point of $p$, using Theorem 3 of Self and Liang
\cite{SL:1987}, it follows that
\be
- 2 (l_{04} - l_{14}) \sim \frac{1}{2} + \frac{1}{2} \chi^2_1.    \label{test-4}
\ee
Here $l_{04}$ and $l_{14}$ denote the maximized log-likelihood values under the null and alternative
hypothesis, respectively.  It may be mentioned that in this case $l_{14} = l_{03}$.

\section{\sc Data Analysis and Simulation Experiments}

\subsection{\sc Data Analysis}

The motivation  for this research comes from a study on altitude decompression sickness conducted by the Air Force Research Laboratory (AFRL).  In the study, human subjects were 
exposed to simulated altitude in a hypobaric chamber. 
Subjects were exposed to different altitudes, varying denitrogenation times, exposure duration, different breathing gas mixtures during the exposure, and levels of exercise as determined by the flight protocol.  During the exposure, subjects were constantly monitored for symptoms of DCS.  If a subject exhibited any of the symptoms associated with DCS, the experiment was terminated and the subject repressurized to ground level.  Subjects who did not exhibit any symptoms during the exposure period are assumed to be censored.  

During exposure to altitude, the gas exchange processes in the tissues are unable to
expel the excess nitrogen causing supersaturation.
These gases which come out of solution when tissues are sufficiently
supersaturated collect as bubbles in the tissue.  To delay the onset of bubbles, researchers suggest either a staged decompression to altitude or prebreathing pure oxygen prior to exposure to allow nitrogen diffusion through desaturation.

The data that we consider here comes from one of the flight profiles that involved a staged ascent.  40 subjects were placed in the hypobaric chamber and exposed to an initial altitude of 18,000 feet.  After four hours (240 minutes), the altitude was increased to 35,000 feet and the experiment continued for another three hours.  In the first four hours of the exposure, prior to the increase in altitude, 7 subjects reported DCS symptoms.  At the end of the study, a total of 31 subjects reported DCS. The remaining 9 subjects were asymptomatic at the end of the study, i.e. they do not show any symptoms of DCS.  In the AFRL study, researchers noted that there were several individuals who remained asymptomatic regardless of the exposure duration and the final altitude.  These individuals represent the cure population, and researchers were interested in determining the characteristics that caused immunity.  The staged ascent profile is an example of a simple step-stress experiment and the presence of a cure population makes the dataset ideal for the proposed model. 

In the experiment, we have $\tau_1=240$ but the value of $\tau_2$ is unknown.  
We fit a cumulative risk model to the data and estimate the unknown parameters for fixed $\tau_2$, assuming it to be known.  We vary the value of $\tau_2$ and employ a discrete optimization with a grid size of 5 minutes to find the value at which the likelihood is maximum.

The MLEs of the parameters of the cumulative risk model assuming Weibull, linear failure rate (LFR) and generalized exponential (GE) hazards, 
their associated standard errors (within parentheses), the 
corresponding maximized log-likelihood (MLL) values, the Kolmogorov-Smirnov (K-S) distances between the empirical and fitted distributions, and 
the $p$ values for the K-S test are presented in Table \ref{Table-no-cov}.

\begin{table}[h]
\caption{Parameter Estimates for the Models without Covariates.  \label{Table-no-cov}}

\begin{center}
\begin{tabular}{|c|c|c|c|c|c|c|c|c|}
\hline  \hline
Model & $\alpha_1$ & $\alpha_2$ & $\lambda_1$ & $\lambda_2$ & $p$  & LL & K-S & $p$-value\\
    &  &  &  &  & &  &  & \\  \hline  \hline
Weibull & 2.4972 & 1.5390 & 0.2341 & 0.9762 & 0.1820 & -38.7093 & 0.0502 & 0.7519 \\
        & (0.4213) & (0.2587) & (0.0378) & (0.1175) & (0.0298) & -- & -- & -- \\
LFR & 0.0985 & 6.8254 & 0.0102  & 1.0915 & 0.2214  & -41.1311  & 0.0949 & 0.2114   \\
        & (0.0118) & (1.2319) & (0.0008) & (0.2478) & (0.0215) & --  & -- & -- \\
GE & 5.0066 & 2.6427 & 1.3461 & 3.6877 & 0.1589 & -39.6485 & 0.0589 & 0.6998  \\
        & (0.9567) & (0.4565) & (0.3534) & (0.7567) & (0.0198) & --  &  -- &  -- \\
KH & 1.2454  & 1.9290 & 0.2003  & 0.7446  & 0.1975  & -39.0482 & 0.1007 & 0.1723  \\
        & (0.1775) & (0.3318) & (0.0397) & (0.1114) & (0.0298) & -- & -- & -- \\
\hline
\end{tabular}
\end{center}
\end{table}
For comparison purposes,  we also fit the Khamis and Higgins \cite{KH:1998}  model with a cure fraction.  The Khamis and Higgins \cite{KH:1998} model with Weibull hazards is a special case of the cumulative risk model  with $\delta$ = 0, i.e. $\tau_1 = \tau_2$

\begin{figure}[t]
\begin{center}
\includegraphics[height=6.0cm,width=7.0cm]{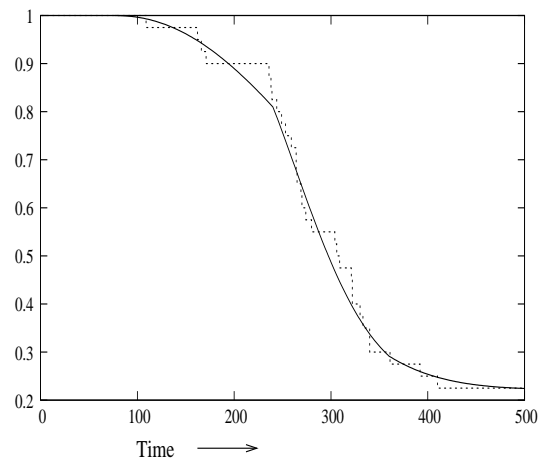}
\caption{Kaplan-Meier estimator and the estimated survival function based on the Weibull model.  Here time is measued in mins.}   \label{pred-1}
\end{center}
\end{figure}

\begin{figure}[t]
\begin{center}
\includegraphics[height=6.0cm,width=7.0cm]{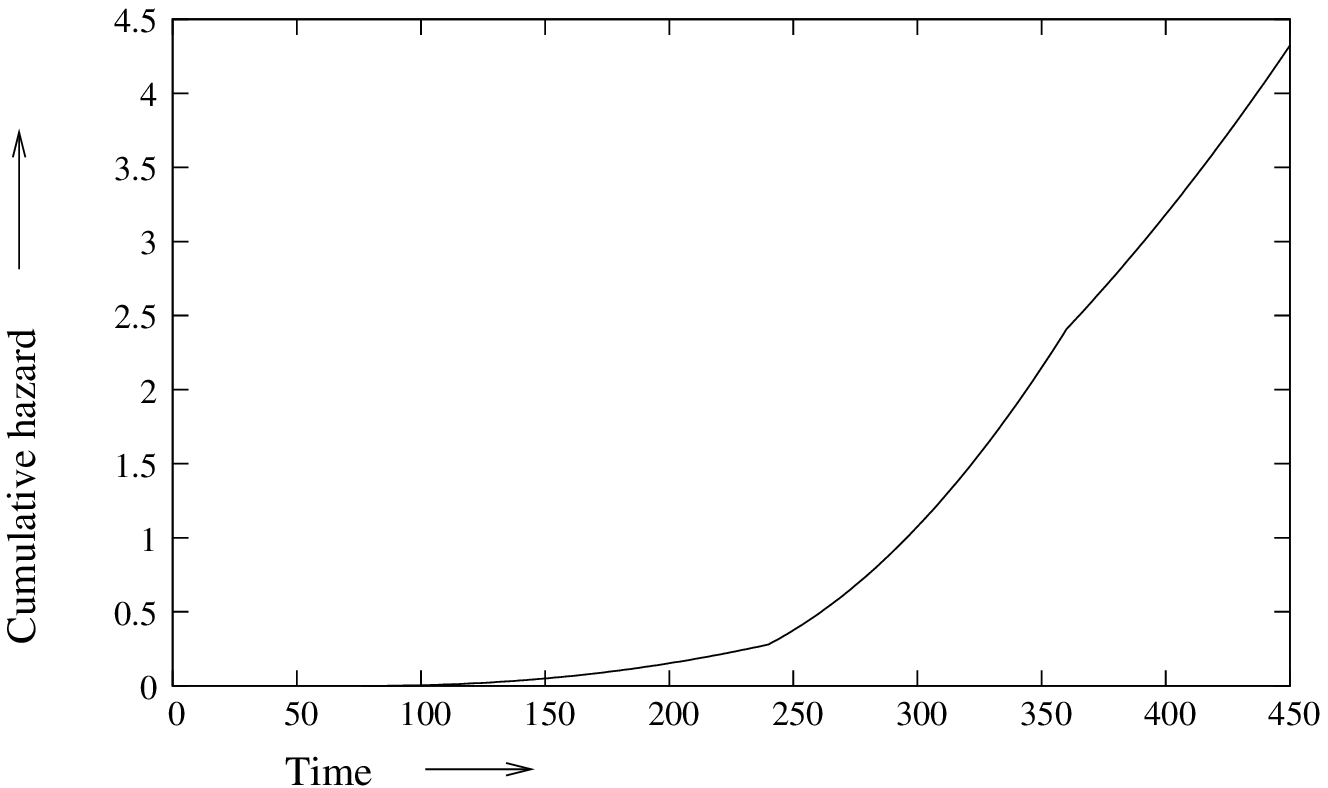}
\caption{Cumulative hazard function of the Weibull model. Here time is measued in mins.}   \label{cumhaz}
\end{center}
\end{figure}

If we examine the maximized log-likelihood values and the K-S distances, it is clear that the Weibull model provides the best fit.  The Kaplan-Meier estimator of the survival function and the survival function based on the Weibull  model are provided in Figure \ref{pred-1}.  The two curves are very close, indicating the Weibull model is indeed a good fit for the data.  
Using a grid search, the estimate of $\tau_2$ was found to be 350 minutes.  This implies a lag of 110 minutes (almost 2 hours) before the effects of the increased altitude are observed.  
The cumulative hazard function with the estimated value of $\tau_2$ is provided in Figure \ref{cumhaz}.  We also obtain an estimate of the cure fraction $p$.  For the Weibull model, we estimate 18\% of the population to be asymptomatic.

We can test whether or not the population has a cure fraction by using the procedures outlined in Section 4.2. 
Under the null hypothesis ${\cal{H}}_{04}$, the MLEs of the unknown parameters and MLL value are
$$
\widehat{\alpha}_1 = 1.9685, \ \ \ \widehat{\alpha}_2 = 1.8170, \ \ \
\widehat{\lambda}_1 = 0.1539, \ \ \  \widehat{\lambda}_2 = 1.0582, \ \ \ MLL = -41.2654.
$$
The value of the test statistic is $ -2(-41.2654 + 38.7093) = 5.3122$. Using the result (\ref{test-4}),
the p-value is  $0.0019$, leading to rejection of the null hypothesis.  This indicates the population has a cure fraction.

In addition to the event times, researchers also collected covariate information on study participants.  For each subject, we have their age (in years), height (in cms), weight (in lbs), body fat percentage (BF) and maximal oxygen uptake ($VO_2$ max in liter/min).  Since the variability in the height and weight measurements was small, we only considered age, body fat and $VO_2$ max in the model.  In Table \ref{Table-2}, we provide information about the minimum (min), maximum (max), median ($Q_2$), first quartile ($Q_1$) and third quartile ($Q_3$)
of the covariates.

\begin{table}[h]
\caption{Summary statistics of the covariates.  \label{Table-2}}

\begin{center}
\begin{tabular}{|c|c|c|c|c|c|}
\hline  \hline
Cov. & min & max & $Q_1$ & $Q_2$ & $Q_3$  \\
    &  &  &  &  &  \\  \hline  \hline
Body fat & 5.9 & 29.9 & 14.6 & 17.1 & 21.1  \\
$VO_2$ max & 1.9 & 6 & 2.7 & 3.2 & 3.95  \\
Age & 21 & 43 & 27 & 34 & 37 \\
\hline
\end{tabular}
\end{center}
\end{table}

Given that the Weibull model provided the best fit for the data without covariates, we assume a Weibull model with covariate vector $Z$ = (1, BF, $VO_2$, Age).  We assume that there is a lag of 110 minutes (i.e. $\tau_2=350$).    
The function $p({\ve \beta}, {\ve z})$ is as follows:
$$
p({\ve \beta}, {\ve z}) = \frac{\exp(\beta_0 + \beta_1*\hbox{BF} + \beta_2*VO_2 + \beta_3*\hbox{Age})}
{1 + \exp(\beta_0 + \beta_1*\hbox{BF} + \beta_2*VO_2 + \beta_3*\hbox{Age})}.
$$
We fit Weibull models with all possible subsets of covariates.  
The maximum likelihood estimates of the unknown parameters and the associated standard errors
(within parentheses) for all the sub-models are provided in Table \ref{Table-cov}.

\begin{sidewaystable}

\centering
\caption{Parameter Estimates for the Weibull Model with Covariates.  \label{Table-cov}}

\begin{tabular}{|c|c|c|c|c|c|c|c|c|c|}
\hline  \hline
Covariates & $\alpha_1$ & $\alpha_2$ & $\lambda_1$ & $\lambda_2$ & $\beta_0$  & $\beta_1$ & $\beta_2$ & $\beta_3$ & MLL \\
    &  &  &  &  & &  &  &  &  \\  \hline  \hline
$z$ = (1,BF, $VO_2$, Age) & 2.6299 & 1.8450 & 0.2215 & 1.1348  & -1.8124 & -3.9196 & 3.6604 & -0.2154 & -31.3009  \\
        & (0.4789) & (0.2918) & (0.0445) & (0.2191) & (0.5412) & (1.1212) & (1.1318) & (0.0519)  & -- \\
$z$ = (1,BF, $VO_2$) & 2.6037 & 1.9657 & 0.2159 & 0.9892  & -1.8503 & -3.9707 & 3.4585 &  & -33.3294  \\
        & (0.4339) & (0.2711) & (0.0401) & (0.1999) & (0.4798) & (1.1198) & (1.1010) & --  &  --\\
$z$ = (1,BF, Age) & 2.5249 & 1.8836 & 0.2339  & 1.0436   & -0.2281  & -3.9895 &  & -1.0607  & -33.6776   \\
        & (0.4412) & (0.2798) & (0.0466) & (0.1814) & (0.0445) & (1.1253) & -- & (0.0498)  & -- \\
$z$ = (1,$VO_2$, Age) & 2.5808 & 1.8008 & 0.2307 & 1.1334  & -2.1468 & --  & 3.9776  & -1.7856 & -34.2477  \\
        & (0.4398) & (0.2810) & (0.0511) & (0.1889) & (0.5798) & -- & (1.2156) & (0.0610)  &  -- \\
$z$ = (1, BF) & 2.4368 & 1.9457 & 0.2269 & 1.0348   & -0.6206  & -3.9817  & --  &  -- & -34.0121   \\
        & (0.4310) & (0.2716) & (0.0515) & (0.1817) & (0.2656) & (1.1191) & -- & --  &  --\\
$z$ = (1, $VO_2$) & 2.5491 & 1.7902 & 0.2228 & 1.1726   & -2.6528  & --  & 3.9823  & --  & -34.5618   \\
        & (0.4397) & (0.2926) & (0.0511) & (0.1898) & (0.6545) & -- & (1.1871) & --   & -- \\
$z$ = (1, Age) & 2.6066  & 1.9207  & 0.2268 & 0.9736   & -0.3374  & -- & -- & -3.0710  & -34.9897   \\
        & (0.4365) & (0.2710) & (0.0516) & (0.1912) & (0.1151) & -- & -- & (1.1097)  & --\\

\hline  
\end{tabular}
\end{sidewaystable}

\begin{figure}[t]
\begin{center}
\includegraphics[height=6.0cm,width=7.0cm]{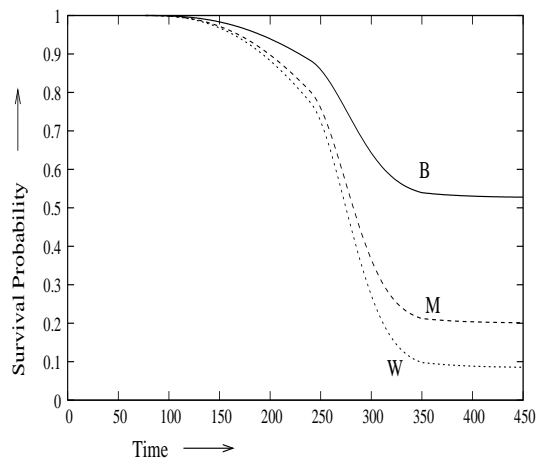}
\caption{Predicted survival function at three different sets of covariates.  Time is measued in mins. Here `B' denotes 
low risk, `M' denots medium risk and `W' denots high risk.}   \label{pred-2}
\end{center}
\end{figure}

To test whether or not the covariates are significant, we use the procedure outlined for Problem 3 in Section 4.2.
The value of the test statistic is -2(31.3099 - 38.7093) = 14.7988,  with a p-value of $0.002$.  We reject the null hypothesis and conclude that the covariates considered here are significant.  As body fat
or age increases, the probability of immunity decreases, and as $VO_2$ increases, the probability of
immunity increases.  These results are consistent with the physiology of decompression sickness.  
Using the Weibull model, we provided the predicted survival function at three different sets of covariates.
(i) Low Risk (B): BF = 5.9, $VO_2$ = 6.0, Age = 21, (ii) Medium Risk (M): BF = 17.1, $VO_2$ = 3.2, Age = 34, (iii)
High Risk (W) BF = 29.9, $VO_2$ = 1.9, Age = 43, in Figure \ref{pred-2}.  We see that the survival functions are quite close up to 240 minutes (the initial exposure level) but the effects of the change in altitude is different for the three groups.  The High Risk group has the lowest proportion of cure individuals.  

\subsection{\sc Simulation Experiments}

In this section we perform some small simulation experiments mainly to see how the proposed methods work for different sample 
sizes and for the above parametric and covariated set up.  We have generated samples from the Weibull model with covariates 
with the following parameters: $\alpha_1$ = 2.6, $\alpha_2$ = 1.8, $\lambda_1$ = 0.22, $\lambda_2$ = 1.14, $\beta_0$ = -1.80, 
$\beta_1$ = -4.0, $\beta_2$ = 3.7, $\beta_3$ = -0.20.  We have considered the lag-period $\delta$ = 100.  We have chosen these
parameters values since they closely mimic the original data.  We have taken diferent sample sizes namely 40, 80, 120, 160 
and 200.  Since in the original data set the BF varies bettwen 5.9 and 29.9, we have chosen at random between these two values
for a generated sample.  Similarly, we have generated at random VO$_2$ between 1.9 and 6.0, and Age between 20 and 43.  We have
calculated the MLEs and the associated mean squared errors (MSEs) based on 1000 replications.  All the results are based on 20\% of 
censored data.  We have reported the average estimates and the square root of the MSEs of all the estimates.  It is observed that as 
sample size increases the MSEs and biases decrease.  In all the cases the EM algorithm stops within 35 iterations.  It seems the 
proposed method works well in this particular set up.

\begin{sidewaystable}

\centering
\caption{MLEs for the Weibull Model with Covariates for Different Sample Sizes.  \label{Table-cov-2}}

\begin{tabular}{|c|c|c|c|c|c|c|c|c|c|}
\hline  \hline
Sample Size & $\alpha_1$ & $\alpha_2$ & $\lambda_1$ & $\lambda_2$ & $\beta_0$  & $\beta_1$ & $\beta_2$ & $\beta_3$ \\
    &  &  &  &  & &  &  &   \\  \hline  \hline
$n$ = 40  & 2.8411 & 2.1105 & 0.2541 & 1.1768  & -2.6816 & -6.1141 & 2.1786 & -0.2576   \\
        & (0.6723) & (0.3871) & (0.0671) & (0.3928) & (0.7129) & (1.3154) & (1.4116) & (0.0723)   \\
$n$ = 80  & 2.7921 & 2.0156 & 0.2487 & 1.1523  & -2.4225 & -5.8791 & 2.6517 & -0.2443   \\
        & (0.3817) & (0.2011) & (0.0401) & (0.1987) & (0.3610) & (0.6578) & (0.7750) & (0.0411)   \\
$n$ = 120  & 2.7015 & 1.9767 & 0.2365 & 1.1511  & -2.1065 & -4.7910 & 2.8917 & -0.2312   \\
        & (0.2218) & (0.1254) & (0.0251) & (0.1310) & (0.2454) & (0.4516) & (0.4519) & (0.0218)   \\
$n$ = 160  & 2.6811 & 1.8917 & 0.2310 & 1.1487  & -1.9167 & -4.4578 & 3.2166 & -0.2189   \\
        & (0.2013) & (0.1045) & (0.0222) & (0.0967) & (0.1810) & (0.3331) & (0.3843) & (0.0202)   \\
$n$ = 200 & 2.6115 & 1.8214 & 0.2242 & 1.1411  & -1.8382 & -4.2331 & 3.5714 & -0.2067   \\
        & (0.1319) & (0.0798) & (0.0156) & (0.0861) & (0.1452) & (0.2512) & (0.2815) & (0.0114)   \\

\hline  
\end{tabular}
\end{sidewaystable}

\section{\sc Conclusions}

In this article,  we propose a new model for step stress experiments that includes a cure fraction.  The susceptible
population has been modeled  using three different parametric forms of the generalized cumulative risk model, and the probability of being
cured is treated as a function of covariates.  Different models are fit to data from a study on altitude decompression sickness, that was the motivation for this research.  The results show that the Weibull cumulative risk model with a cure fraction provides  
an excellent fit to the data.  Using the model, we were able to determine the significant covariates, estimate the cure proportion, and find an estimate of the lag period. The model is extremely flexible and can be easily adapted to different distributions and censoring schemes.

\section*{\sc Acknowledgements:} The authors would like to thank the reviewers for their
constructive suggestions which have helped to improve the manuscript significantly.

\end{document}